\magnification = 1200
\hsize=6truein
\hoffset=0.25truein
\parskip=\smallskipamount
\def\rd#1{\mathop{{\rm d}#1}\nolimits}

\centerline{\bf Fractional Electrons in Liquid Helium?}
\bigskip
\centerline{R. Jackiw}
\centerline{{\it Massachusetts Institute of Technology, Cambridge, MA
02139\/}}
\smallskip
\centerline{C. Rebbi}
\centerline{{\it Boston University, Boston, MA 02215\/}}
\smallskip
\centerline{J.R.
Schrieffer}
\centerline{{\it Florida State University, Tallahassee, FL 32310\/}}
\medskip
\centerline{13 December  2000\qquad MIT-CTP\#3059}
\bigskip
\centerline{Abstract}

{\narrower\noindent 
We argue that electrons in liquid helium bubbles are not
fractional, they are entangled.\par}
\bigskip

\noindent
In an analysis of several unexplained observations on the mobility of
electron-inhabited ``bubbles'' in liquid helium, H.~Maris has suggested
that such bubbles can fission into two smaller daughter bubbles, each
containing half of the original electron's wave function and each allowing
the detection of fragments of the original electron, which has divided
into two pieces that act as though they are fractions of the original
particle~[1].

In this connection it is important to keep in mind two significantly
different concepts of fractional fermions.

On the one hand, there are physical situations where each individual
measurement yields a fractional result.  Two examples are the field-theoretic
models for fermions propagating across domain walls, as realized
experimentally by solitons on polyacetylene~[2,3]; and also the excitations in
the fractional quantum Hall effect~[4].  For these the fractional characteristics
are sharp observables (quantum mechanical eigenvalues) without
dispersion~[5], and the phenomena put into evidence previously unsuspected
fermion fractionization.

Alternatively, the fraction is an expected value, not an eigenvalue.
Then repeated measurements always yield either the full quantum numbers
of the particle or a null result, and the ``fraction'' is just
the probability of finding the full particle.  This
is characterized by a nonvanishing dispersion, which remains nonzero
no matter how far apart the measurements are performed.

The phenomenon discussed by Maris belongs to the second class:
fractional quantum numbers do not arise, only fractional expectations
with nonvanishing dispersion.  The reason for this is the following.
Let the electron's wave function be presented as $\psi_+=(\psi_1+\psi_2)/
\sqrt{2}$, where $\psi_1$ and $\psi_2$ are (normalized) wave
functions peaked at the first and
second daughter bubbles, respectively.  The expected electron number
localized around the first bubble is
$$
n=\langle + \vert N_f \vert + \rangle = \int \rd V f \psi_+^* \psi_+\ .
$$
Here $f$ is a sampling function, which localizes the volume integral in the
region of the first bubble.  It is true that $n = 1/2$.

However, one must also look at the variance $(\Delta n)^2=
\langle + \vert N_f^2 \vert + \rangle - n^2$.
We remember that there exists another state  $\psi_-=(\psi_1-\psi_2)/
\sqrt{2}$ that is almost degenerate with the state $\psi_+$. Retaining just
these two states, we find:
$$\displaylines{
(\Delta n)^2=
\langle + \vert N_f \vert + \rangle\langle + \vert N_f \vert + \rangle
+\langle + \vert N_f \vert - \rangle\langle - \vert N_f \vert + \rangle
-n^2 = \bigl\vert \langle + \vert N_f \vert - \rangle \bigr\vert^2
\cr\noalign{\medskip}
= \Bigl| \int \rd V f \psi_+^* \psi_- \Bigr|^2 = {\textstyle{1 \over 4}}\ .
}
$$
Thus no matter how far apart the two bubbles are taken, one cannot
isolate a sharp fraction.

Indeed the effect is a standard quantum mechanical result for an
``entangled'' state.  Here the entanglement is in location:  the electron
is either in one bubble or the other, but before the measurement one
cannot decide in which bubble it resides.  While ``half the electron's wave
function'' is in both bubbles, measurements will find a full electron in half the
bubbles.

Finally, we observe that (as remarked by K.~Canter) inasmuch as the
helium bubble is stabilized by  Pauli repulsion between the electron in the
bubble and the orbital electrons in helium, for a bubble to exist its inhabitant
must be identical with the electrons in the atoms, and cannot be just a fraction
thereof.
\bigskip

\frenchspacing 

\item{[1]} H. Maris, {\it J. Low Temp. Phys.}, {\bf120}, 173 (2000); see also
{\it Physics Today}, {\bf53}, No.~11, 9 (2000) and {\it New Scientist},
No.~2260, 25 (2000).

\item{[2]} R. Jackiw and C. Rebbi, {\it Phys. Rev. D} {\bf13}, 3398 (1976). 

\item{[3]} W.-P. Su, J.R. Schrieffer, and A. Heeger, {\it Phys. Rev. Lett.}
{\bf42}, 1698 (1979); R.~Jackiw and  J.R. Schrieffer, {\it Nucl. Phys.} {\bf
B190} [FS3],  253 (1981).

\item{[4]} R.B. Laughlin, H.L.~Stormer, and D.C. Tsui, {\it Rev. Mod. Phys.}
{\bf71}, 863 (1999).

\item{[5]} S. Kivelson and J.R. Schrieffer, {\it Phys. Rev. B} {\bf25}, 6447
(1982); R.~Rajaraman  and J.S.~Bell, {\it Phys. Lett.} {\bf116B}, 151 (1982); 
R.~Jackiw, A.~Kerman, I.~Klebanov, and G.~Semenoff, {\it Nucl. Phys.} {\bf
B225} [FS9],  233 (1983).

\bye